\documentclass[12pt, preprint]{aastex}
\usepackage{epsfig}

\begin{document}

\title{Concerning the existence of a `turbulent' magnetic field
in the quiet Sun}

\shorttitle{Turbulent magnetic fields in the quiet Sun}

\author{Rafael Manso Sainz\altaffilmark{1,2}, 
Egidio Landi Degl'Innocenti\altaffilmark{3} \& Javier Trujillo Bueno\altaffilmark{2,5}}
\altaffiltext{1}{High Altitude Observatory, National Center for Atmospheric Research\altaffilmark{4}, P. O. Box 3000,
Boulder CO 80307-3000}
\altaffiltext{2}{Instituto de Astrof\'{\i}sica de Canarias, E-38205, La Laguna,
Tenerife, Spain}
\altaffiltext{3}{Dipartimento di Astronomia e Scienza dello Spazio, Largo Enrico Fermi 2,
I-50125, Firenze, Italy.}
\altaffiltext{4}{The National Center for Atmospheric Research is sponsored by the National Science Foundation}
\altaffiltext{5}{Consejo Superior de Investigaciones Cient\'{\i}ficas (Spain)}


\begin{abstract}

	We report on the $a\,{}^5\!F$-$y\,{}^5\!F^\circ$ 
	multiplet of Ti {\sc i} and its interest for the study of 
	``turbulent'' magnetic fields in the quiet solar photosphere.
	In particular, we argue that the sizable scattering polarization
	signal    of the 4536 \AA\ line (whose lower and upper levels
	have Land\'e factors equal to zero),
	relative to the rest of the lines in the multiplet, gives
	{\em direct} evidence for the existence of 
	a ubiquitous, unresolved magnetic field.
	We cannot determine precisely the strength
	of the magnetic field,
	but its very existence is evidenced 
	by the differential Hanle effect technique that this 
	Ti {\sc i} multiplet provides.
\end{abstract}


Magnetic fields in the solar atmosphere
may be detected most clearly by the 
circular polarization pattern they induce in spectral lines. 
However, this Zeeman effect technique is not very suitable 
for investigating magnetic fields that
have complex unresolved geometries because the
contributions of opposite magnetic polarities within the spatio-temporal resolution element of the observation tend to cancel out.
Consequently, extremely weak magnetic fields or even strong fields with a rather convoluted topology may remain virtually undetectable.

This is unfortunate since, according to our general picture of
solar magnetism, most regions
of the solar atmosphere are expected to be filled by magnetic fields
whose geometries are complex and far from 
being spatially resolved.
In particular, `turbulent' magnetic fields driven by highly chaotic fluid motions are supposed to 
pervade the `quiet' regions of the solar photosphere, with mixed magnetic polarities even below the 
photon mean-free-path (e.g., Cattaneo 1999).
At higher chromospheric and coronal levels the medium rarefies exponentially,
the magnetic fields weaken and the picture becomes still more problematic.

The presence of relatively weak magnetic fields may be revealed by
the modification they produce in the linear polarization pattern
generated by scattering processes 
in spectral lines (Hanle effect).
Typically (but not always), the change
consists in a net depolarization and a rotation of the polarization plane.
If the azimuth of the
magnetic field has a random variation 
within the observed resolution element, rotations
of polarization planes cancel out, but the reduction of the
scattering polarization amplitude remains. Therefore, the
Hanle effect has the potential for detecting the presence of
tangled magnetic fields on subresolution scales 
in the solar atmosphere (Stenflo 1982).

It seems then natural to rely on this technique to probe the kind of 
`microturbulent' photospheric magnetic fields discussed above.
This has been tried out by some authors in the past, concluding 
that depolarization by a 'turbulent' field is required to be able 
to explain the available observations of scattering polarization 
on the sun (Stenflo 1982; Faurobert-Scholl et al.
1995, 2001; Stenflo et al. 1998).
The main problem of these analyses is that they require a reference
zero-field polarization amplitude which, in turn, heavily relies
on theoretical calculations. 
Only very recently such radiative transfer calculations have started 
to be sufficiently realistic, taking into account the three-dimensional 
and dynamic nature of the solar atmosphere (Trujillo Bueno, Shchukina 
\& Asensio Ramos 2004).
In any case, the question may always be raised of whether all the other 
relevant physical mechanisms that can produce depolarization have been 
accounted for and whether the observed depolarization can then be safely 
ascribed to the presence of a magnetic field. 
In order to achieve model independence it is useful to consider various
spectral lines with different sensitivities to the Hanle effect 
(Stenflo et al. 1998).
Unfortunately, the particular combination of atomic lines used by
Stenflo et al. (1998) is far from optimum, because they have different
line formation properties and none of them is insensitive to magnetic fields.
Here we present an analysis of the polarization pattern of the 
multiplet 42 of Ti {\sc i}.
We argue that the available observations of scattering polarization
in this multiplet provide direct evidence that magnetic depolarization 
due to a `turbulent' field is really
at work in the quiet solar photosphere.

Recently, Manso Sainz \& Landi Degl'Innocenti (2003a,b; 
hereafter Papers {\sc i} and {\sc ii}, respectively) have investigated
the physical origin of the observed linearly-polarized, 
solar limb spectrum of Ti {\sc i}.
To this end, these authors used
a very realistic multilevel model for 
the Ti {\sc i} atom, which includes a very interesting
multiplet that may be of great relevance for the study of solar magnetism
via the Hanle effect.
This multiplet is formed by the transitions between the 
lowest term of the titanium's quintet system ($a\,{}^5\!F$) 
and the excited $y\,{}^5\!F^\circ$ term (see Fig.~1).
Interestingly, all the lines in the multiplet lie in the blue part
of the visible spectrum and their scattering polarization signals have, 
in fact, been recently observed in quiet regions near the solar
limb (Gandorfer 2002).
The important thing about this multiplet is that the transition
at 4536 \AA\ shows
a large polarization peak relative to the other lines of the multiplet
and that the Land\'e factor of both its lower and upper level 
($a\,{}^5\!F_1$ and $y\,{}^5\!F^\circ_1$, respectively) is zero.
In other words, the line at 4536 \AA\ is,
unlike the others of the same multiplet, essentially insensitive
to magnetic fields.
We show here that the most natural explanation 
of the reported linear polarization observations invokes the existence
of a tangled magnetic field permeating the solar atmosphere 
within the resolution element.

Panel (a) of Fig.~2 shows the observed fractional linear polarization 
in the $a\,{}^5\!F$-$y\,{}^5\!F^\circ$ multiplet as reported by
Gandorfer (2002).
The transition at 4527.3~\AA\ will not be considered further in our 
analysis for its being blended with two other transitions of 
Ce {\sc ii} and Cr {\sc i}, two species seemingly quite `active' 
in scattering line polarization.
The first thing to note is that all these lines are strong
in the sense of Paper {\sc ii}, i.e., they have residual intensities
$(I-I_c)/I_c$ larger than 0.5, and all of them, with the notable exception of $\lambda$4536, have $Q/I$ signals below the continuum polarization level.
In the presence of a significantly polarized continuum 
the contribution of the strongest spectral lines 
to the linearly polarized spectrum cannot be simply added up
and the depolarization of the continuum by the presence of 
the spectral line cannot be neglected.
Nevertheless, we can always assume a similar effect for 
all the lines in the multiplet and compare the $Q/I$ signals
between them, in particular, with the 4536~\AA\  line.

A theoretical estimate for the $Q/I$ structure of the multiplet is
shown in panel (b) of Fig.~2.
It has been calculated for a 90$^\circ$ observation of a layer of
atomic titanium gas which is illuminated from below by a photospheric-like
radiation field. Details on the atomic model and radiation field
are given in Paper {\sc i}. Firstly,
we have calculated the atomic polarization
that optical pumping processes induce
in all the levels of the atomic model in the absence of any
depolarizing mechanism. Secondly, we have calculated
the $Q/I$ signal in each transition taking into
account the emissivity as well as dichroism effects 
(see Eq.~(14) of Paper {\sc i}).
As demonstrated in Paper {\sc i}, this method is able
to reproduce (at least qualitatively) the structure of the 
whole second solar spectrum of Ti {\sc i}.

In order to model the depolarization 
due to a weak microturbulent magnetic field we assume one and the same 
depolarizing rate for all the atomic levels, except for
the levels $a\,{}^5\!F_1$ and
$y\,{}^5\!F^\circ_1$ which have zero Land\'e factor (Sugar \& Corliss 
1985)\footnote{See {\tt http://physics.nist.gov/}}. 
We point out that all the ${}^5\!F$ terms of titanium 
are well described by the LS-coupling. Therefore,
according to the well known formula for the Land\'e 
factor we can assume that all the ${}^5\!F_1$ and ${}^5\!F^\circ_1$ 
levels are completely insensitive to the magnetic field. 
Panels (c) and (d) show the $Q/I$ structure of the 
$a\,{}^5\!F$-$y\,{}^5\!F^\circ$ multiplet assuming depolarizing rates $D$
of $10^8$ s$^{-1}$ and $5{\times}10^8$ s$^{-1}$, respectively, for all
except the ${}^5\!F_1$ and ${}^5\!F^\circ_1$ levels (thick lines).
If we roughly estimate an `equivalent' magnetic field intensity
through the equation
$2\pi\nu_L g=8.79{\times}10^6\,B\,g\approx D$ 
($\nu_L$ and $g$ being the Larmor frequency
and Land\'e factor, respectively), then, taking $g{\approx}1$,
a depolarizing rate $D=10^8$ s$^{-1}$ 
corresponds to a magnetic field of the order of 12 gauss, while 
$D=5{\times}10^8$ s$^{-1}$ implies 60 gauss.
For comparison, dotted lines show the $Q/I$ values when all 
the levels are depolarized without exception.
Only the transitions involving at least one level
with zero Land\'e factor
($\lambda\lambda$4527, 4536 and 4544) change their behavior,
as expected.

It is of interest to note that this $D=5{\times}10^8$ s$^{-1}$ 
case produces a relatively good agreement with 
the $Q/I$ observations of Gandorfer (2002), 
and that Trujillo Bueno et al. (2004) concluded that
when one assumes a volume-filling and single-valued microturbulent
field then the best theoretical fit
to the observed scattering polarization in the Sr {\sc i} 4607 \AA\ line
is obtained for $B_{\rm microturbulent}{\approx}60$ gauss.

Since we are considering transitions belonging to the same multiplet,
we can assume that they `form' in the same atmospheric region and hence,
under similar environmental conditions.
If this formation region were
essentially devoid of magnetic fields
(say, a significant volume filled with magnetic fields below one gauss), then we would expect relative $Q/I$ signals qualitatively resembling
the solid line in Fig.~2b or the dotted lines in Figs.~2c, d.
In particular, the ${\Delta}J=0$ transitions 
$\lambda\lambda$4533, 4534 and 4535 would show $Q/I$ signals similar 
or even larger than the $\lambda$4536 transition.
However, the observed large polarization peak in the 4536 \AA\ line 
relative to the others, can only be explained if the depolarization
is of magnetic origin and therefore does not affect the ${}^5\!F_1$ levels.

Finally, we would like to point out that the main deficiency of our analysis
is our neglecting of the continuum polarization.
At such blue wavelengths the continuum polarization due to Rayleigh and/or Thomson
scattering is already significant, and lines as strong as the ones
in this 
multiplet show an important interaction with it 
that cannot be neglected (see discussion in Paper {\sc ii}).
For this reason we have presented our theoretical calculations 
in Fig.~2 without superimposing the continuum.
These considerations should be taken into account in further,
more quantitative analyses.
Yet, they do not change our conclusions.

Summarizing: 
among the several observed solar lines of the $a\,{}^5\!F$-$y\,{}^5\!F^\circ$ multiplet of titanium, only the line $\lambda$4536 shows a 
polarization peak, all others showing just a continuum depolarization.
Since the only apparent peculiarity of
the 4536\AA\ line lies in its lower and upper levels having zero 
Land\'e factor (i.e., its insensitivity to magnetic fields),
the observed polarization pattern may be considered 
as direct evidence for
the presence of a tangled magnetic field in the solar
atmosphere, which is depolarizing  the scattering polarization signals
of all the lines in the multiplet except, 
obviously, for $\lambda$4536.

The arguments above rely on a direct interpretation 
of observations and on a relative independence of theoretical modeling
(actually, this is where their strength lies).
However, since we are not solving the radiative transfer problem
we cannot account for the details of the polarization patterns
arising from the different illumination and depolarizing conditions
at the height of formation of the individual lines. 
This prevents, in general, our comparing polarization signals 
between spectral lines belonging to different multiplets
to obtain quantitative estimates of the magnetic field.
Notwithstanding, we can compare lines belonging 
to the same multiplet assuming that they form under 
similar thermodynamical, radiative and magnetic conditions.
We can then make qualitative interpretations of the most conspicuous 
polarization patterns.
That has been precisely the case here.

In this respect, it must be noticed the interesting multiplet 
$a{}^3\!H$-$x{}^3\!H$, whose line at 4742.8 \AA\ 
shows a very conspicuous polarization signal (actually one of the strongest
of the whole second solar spectrum of Ti {\sc i}), which is twice
as large as the signal observed in 
the $\lambda\lambda$4758.1 and 4759.3 lines of the multiplet.
This cannot be easily explained in terms of the simple modeling 
of Paper {\sc i} (all three would show large, but similar 
amplitudes), 
nor by some kind of 'differential' action of the magnetic
field, since all upper levels of the three transitions have similar
critical fields (the field at which the Zeeman splitting is of the order
of the natural width of the level): $\approx$ 11, 9 and 8 gauss, respectively.
This anomalous behavior is probably related to an interesting spectroscopic
feature of the lower level of $\lambda$4742.8.
Contrary to the other levels involved in the multiplet, this level 
(with total angular momentum $J=4$) is not 
well described as a purely LS-coupling level.
Actually, it is a 52\% -- 44\% mixing of the two configurations
$3d^3({}^2\!H)4s \, {}^3\!H$ and $3d^3({}^2\!G)4s\, {}^1\!G$, respectively
(see Sugar \& Corliss 1985).
(It is interesting to note that 
the lower level ($b{}^1\!G$) of the transition  
showing the largest observed $Q/I$ signal of
the titanium's second solar spectrum ($\lambda$5644.1) 
is a 48\% -- 48\% mixing
of the very same configurations $3d^3({}^2\!H)4s \, {}^3\!H$ and 
$3d^3({}^2\!G)4s\, {}^1\!G$, respectively.)
Due to this accidental coincidence the lower and upper levels of 
$\lambda$4742.8 are spectroscopically similar to an equivalent
LS-coupling level (Land\'e factor, total deexcitation probability,
and critical field), but the line itself has a smaller transition 
probability relative to $\lambda\lambda$4758.1 and 4759.3. 
Consequently,  the line  $\lambda$4742.8 probably forms 
at slightly different 
height in the photosphere than the other two, where the 
radiation field may be dissimilar.

We think that, all in all, the interpretation of the polarization
pattern in the  $a\,{}^5\!F$-$y\,{}^5\!F^\circ$ multiplet
of neutral titanium given in this letter is direct 
evidence for the existence of an unresolved 
magnetic field permeating large volumes of the 
quiet solar atmosphere. 
In any case, further study of the formation of the second solar spectrum
of Ti {\sc i} is necessary
in order to  arrive to more definite
conclusions.
By combining the Hanle-effect
line ratio technique proposed here for the relatively strong
13 lines of this Ti {\sc i} multiplet with that reported by Trujillo Bueno (2003)
for the weak ${\rm C}_2$ lines of the Swan system,
we can now aim at further exploring
the full complexity of the small-scale magnetic activity
of the `quiet' Sun. To this end, we have already initiated
systematic observations of both atomic and molecular lines as well as 
more detailed theoretical investigations.

\acknowledgments
This research has been partly funded by the
Spanish Ministerio de Educaci\'on y Ciencia
through project AYA2001-1649 and by the European
Solar Magnetism Network. We thank 
Roberto Casini for several interesting discussions on
atomic spectroscopy and Jan Olof Stenflo for suggesting improvements
to the original version of this letter.

\clearpage
\begin{figure}
\centering
\includegraphics[width=7cm]{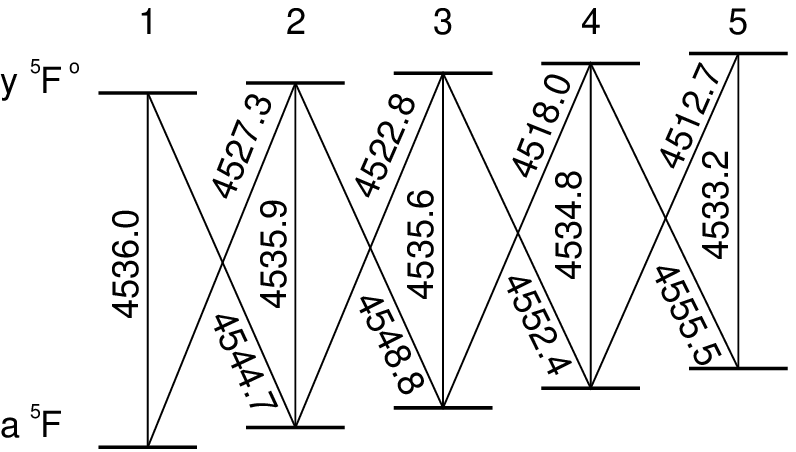}
\caption{Transitions (wavelengths in \AA) 
	between the $a\,{}^5\!F$ and $y\,{}^5\!F^\circ$ terms of Ti {\sc i}.
	Upper labels indicate the total angular momentum of the levels.}
\end{figure}

\clearpage
\begin{figure}
\centering
\includegraphics[width=8cm, bb=0 110 255 595]{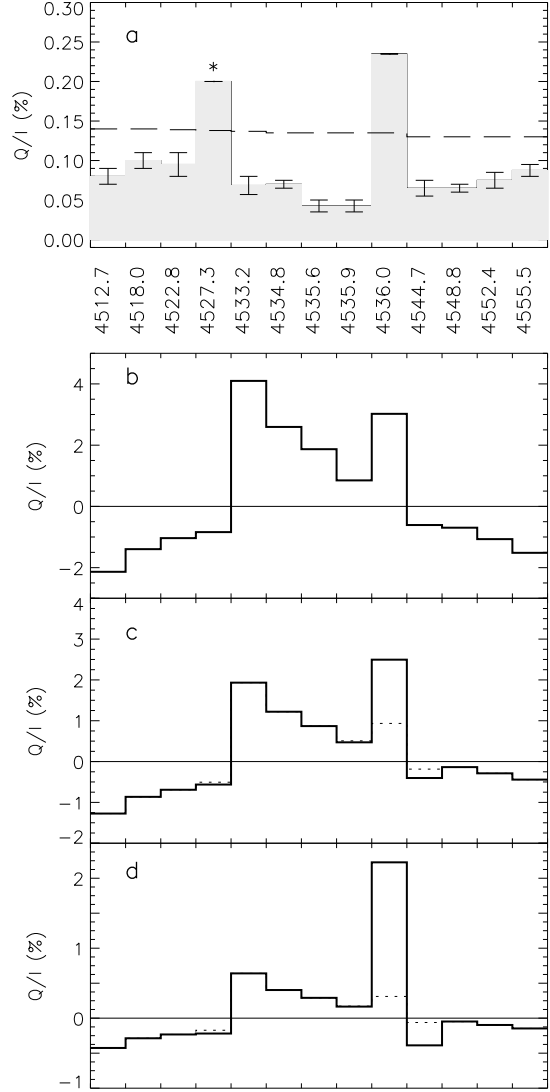}
\caption{Fractional polarization $Q/I$ in the thirteen spectral lines 
	allowed between the $a\,{}^5\!F$ and $y\,{}^5\!F^\circ$ 
	terms of Ti {\sc i}.
	{\bf a)} observations near the solar limb as reported 
	in Gandorfer (2002).
	{\bf b)} theoretical $Q/I$ calculated as explained in the text.
	{\bf c)-d)} as panel {\bf b)} assuming depolarizing rates
	10$^8$ s$^{-1}$ and $5{\times}10^8$ s$^{-1}$, respectively, for all
	Ti {\sc i} levels except the $a{\,}^5\!F_1$ and 
	$y{\,}^5\!F^\circ_1$ones. 
	Note the change of the $Q/I$ scale between panels.
	The meaning of the dotted lines is explained in the text, while
	the symbol * indicates a Ti {\sc i} line that we do not consider
	in our analysis because heavily blended.}
\end{figure}

\end{document}